\documentstyle[11pt,
		epsfig]{article}
\newskip\humongous \humongous=0pt plus 1000pt minus 1000pt
\def\caja{\mathsurround=0pt} \def\eqalign#1{\,\vcenter{\openup1\jot
\caja   \ialign{\strut \hfil$\displaystyle{##}$&$
\displaystyle{{}##}$\hfil\crcr#1\crcr}}\,} \newif\ifdtup

\def\frac#1#2{ {{#1} \over {#2} }}

\def\ltap{\raisebox{-.4ex}{\rlap{$\sim$}} \raisebox{.4ex}{$<$}}
\def\gtap{\raisebox{-.4ex}{\rlap{$\sim$}} \raisebox{.4ex}{$>$}}

\def\ie{\hbox{\rm i.e. }}

\def\kir{k_{{\rm ir}} }
\def\kuv{k_{{\rm uv}} }
\def\lsd{\ell_{{\rm sd}} }

\def\s{\sigma}
\def\d{\delta}

\def\al{\alpha}
\def\as{\alpha_S}
\def\beq{\begin{equation}}
\def\eeq{\end{equation}}
\def\bit{\begin{itemize}}
\def\eit{\end{itemize}}
\def\re#1{(\ref{#1})}

\def\np#1#2#3{Nucl.\ Phys.\ B#1 (19#3) #2}
\def\pl#1#2#3{Phys.\ Lett.\ #1B (19#3) #2}

\def\prep#1#2#3{Phys.\ Rep.\ #1 (19#3) #2}

\begin{document}
\begin{titlepage}
\begin{flushright}
     UPRF 96-488\\
     IFUM-565-FT\\
     November 1996 \\
\end{flushright}
\par \vskip 10mm
\begin{center}
{\Large \bf
Infrared renormalons and finite volume
\footnote{Research supported in part by MURST, Italy
and by EC Programme ``Human Capital and Mobility",
contract CHRX-CT92-0051 and contract CHRX-CT93-0357.}}
\end{center}
\par \vskip 2mm
\begin{center}
F.\ Di Renzo$\,^a$,
G.\ Marchesini$\,^b$
and  E.\ Onofri$\,^c$ \\
\vskip 5 mm
$^a\,${\it Department of Mathematical Sciences, \\
University of Liverpool, United Kingdom}\\
\vskip 2 mm
$^b\,${\it Dipartimento di Fisica, Universit\`a di Milano \\
and INFN, Sezione di Milano, Italy}\\
\vskip 2 mm
$^c\,${\it Dipartimento di Fisica, Universit\`a di Parma \\
and INFN, Gruppo Collegato di Parma, Italy}
\end{center}
\par \vskip 2mm
\begin{center} {\large \bf Abstract} \end{center}
\begin{quote}
We analyze the perturbative expansion of a condensate
in the $O(N)$ non-linear sigma model for large $N$ on a two dimensional
finite lattice. On an infinite volume this expansion is affected by
an infrared renormalon.
We extrapolate this analysis to the case of the gluon condensate of
Yang-Mills theory and argue that infrared renormalons can be detected
by performing perturbative studies even on relatively small lattices.
\end{quote}
\end{titlepage}

\section{Introduction}

In $SU(3)$ Yang-Mills theory on the lattice, the
coefficients of the perturbative expansion of the simplest
observables, the elementary Wilson plaquettes, have been
computed to large order \cite{Pisa,DMO} and they show evidence
of factorial growth.
These perturbative calculations are needed not only to clarify the
presence and relevance of infrared (IR) renormalons \cite{IRren} but
also to extract the values of physical quantities, such as heavy quark
decay constants \cite{HQ} from the results of Monte Carlo simulations.

The perturbative coefficients of the elementary Wilson plaquette,
an observable corresponding to the $SU(3)$ gluon condensate, are
known analytically \cite{Pisa} up to the third order in $\as$,
the weak-coupling parameter at the ultraviolet lattice scale,
related to the usual lattice constant $\beta$ by
$4\pi \as= 6/\beta$.
At larger orders these coefficients have been obtained \cite{DMO}
by a numerical method based on the stochastic formulation
of the theory \`a la Parisi--Wu \cite{PW}.
One considers the weak coupling expansion of the Langevin
equations and solves numerically the truncated set of equations
corresponding to a given perturbative order.
The coefficients have been computed numerically up to the order
$\as^8$ on a $8^4$ lattice and up to the order $\as^6$ on a $12^4$
lattice.
In the sequel we shall always consider symmetric 
square (d=2) or hypercubic (d=4) lattices with
$M$ lattice points in each direction.

Up to the sixth order the differences between the coefficients for
$M=8$ and $M=12$ are found to be contained
within the statistical errors, at present smaller than $10\%$
\footnote{ Since the stochastic method requires the
introduction of two copies of the field for each order in $\as$, the
amount of computer memory grows proportionally to $ M^4 \times $
number of loops. This is why only rather small lattices have been
considered so far.}.

An important question in lattice calculation is to what extent the
finite volume affects the results.
In particular, if the perturbative part of an observable is affected
by an IR renormalon, the factorial behaviour of its coefficients is
tamed by the finite volume.
Then we should face the question\footnote{
This problem has been stressed to us by Martin Beneke.}
whether the lattice size ($M=8,12$) considered in
\cite{DMO} are sufficiently large to detect the IR renormalon
factorial growth.

In order to have a simple estimation of the finite volume effect
we consider the typical integral involving an IR renormalon
and we introduce an infrared cutoff related to the finite volume.
For a condensate with dimension $\d$ we consider
\beq\label{WM}
W_\d(\as,M) \equiv
\int_{\kir^2}^{\kuv^2}\frac{dk^2}{k^2}
\left(\frac{k}{\kuv} \right)^\d\;\as(k^2)
\,, \;\;\;\;\;
\kuv=\frac{\pi}{a}
\,, \;\;\;\;\;
\kir=\frac{2\pi}{Ma}
\,,
\eeq
with $\kuv$ and $\kir$ the infrared and ultraviolet cutoff
on the lattice of size $M$ and spacing $a$.
Using the one loop form of the running coupling
\[
\as(k^2)=\frac{\as}{1+b_0\,\as\,\ln(k^2/\kuv^2)}
\,,
\]
one obtains the formal expansion
\beq\label{usual}
W_\d(\as, M)=\sum_{\ell=1} \as^{\ell}\; c_\ell(M)\,,
\;\;\;\;\;
c_\ell(M)=
b_0^{\ell-1}
\;\int_{\kir^2}^{\kuv^2}\frac{dk^2}{k^2}
\left( \frac{k}{\kuv} \right)^\d \;
\; \ln^{\ell-1} \left( \frac{\kuv^2}{k^2}\right)
\,.
\eeq
The factorial growth of the coefficients is obtained from values
of $k$ around the steepest descent point
\beq
k_{{\rm sd}} = \kuv\; e^{-(\ell-1) / \d}
\,.
\eeq
The IR renormalon factorial growth is then obtained only if
$k_{{\rm sd}} > \kir$, \ie the steepest descent point lies inside the
integration region, which gives the upper bound
\beq\label{ellsd}
\ell < \lsd = \d \; \ln\left( \frac{M}{2} \right)+1
\,.
\eeq
Above this value the factorial growth is tamed. For $M=8,\;12$ and
$\d=4$ one obtains $ \ell_{{\rm sd}} \simeq 6.5, \;8.2$. Therefore,
among the coefficients computed in \cite{DMO} only the ones with
$\ell \le 6$ would be inside the bound. For higher order
coefficients one should use lattices larger than $M=8$.

In this paper we try to probe the reliability of the indication
\re{ellsd} and we shall argue that the factorial growth of the
coefficients is still present at  values of $\ell$ much larger
than $ \ell_{{\rm sd}}$.

First we consider the exactly solvable asymptotically free case
given by the large $N$ limit of the $O(N)$ non-linear sigma model in
a two dimensional finite lattice. We consider the perturbative
expansion of the condensate with $\d=2$ and we compare the
exact coefficients and their behaviour in $M$ with the ones given
by \re{usual}.
We show that the large order coefficients agree with the ones
obtained from the typical integral \re{WM} provided that 
the running coupling $\as(k^2)$ in the integrand is replaced by
$\as(r k^2)$ with $r <1$.
We then apply this idea to the four dimensional case and consider
the expression in \re{WM} with $\d=4$.
As suggested by the non-linear sigma model, we assume that the
running coupling is taken at a scale smaller than $k^2$.
We then argue that the factorial growth extends to larger values of
$\ell$ well inside the values considered in \cite{DMO}.

\section{Non-linear $O(N)$ sigma model at large $N$}

IR renormalons are present in the perturbative expansion of any
observable in the non-linear sigma model in two dimensions
\cite{ON}.
We consider the lattice version of the model \cite{CR} on a finite volume.
The internal energy density $U$, which is a condensate of dimension
$\d=2$, has the following operator product expansion
\beq\label{OPE}
U=a^{-2}U_{{\rm pert}} + m^2U_1 +m^4U_2 + \,  \cdots
\,,
\eeq
with $a$ the lattice UV cutoff and $m$ the scale parameter generated
by the breaking of  scale invariance. In the limit $N\to\infty$ one has
\beq\label{scale}
(am)^2 = 32\; e^{-4\pi\beta}
\,,
\eeq
where $\beta$ is the strong-coupling parameter at the lattice UV scale.
The perturbative component has the following large $N$ expansion
\beq
U_{{\rm pert}} =\frac{N-2}{2N}\,4\pi\al
\;+\; \frac{1}{2N}\,W(\al) \;+\; O(N^{-2})
\,,
\eeq
where we have defined the weak-coupling by $4\pi\al = 1/\beta$.

The first $1/N$ correction contains IR renormalons.
For finite volume, with $M$ lattice size, and for infinite
volume one finds
\beq\label{W2}
\eqalign{
&W(\al, M)=
\frac{1}{M^2}\sum_{n_1,n_2}
\;\; \frac{4\pi\al}{1+4\pi\al A_M(k_1,k_2)}\,,
\cr&
W(\al,\infty)=
\int_{-\pi/a}^{\pi/a}\frac{a^2d^2k}{(2\pi)^2}
\;\; \frac{4\pi\al}{1+4\pi\al A_\infty (k_1,k_2)} \,,
}
\eeq
with
\beq\label{A}
\eqalign{
&
A_M(k_1,k_2)=
\frac{1}{2M^2}\sum_{m_1,m_2}
\;\; \frac{\hat{k}^2-\hat{p}^2
-(\widehat{p+k})^2}{\hat{p}^2(\widehat{p+k})^2}\,,
\cr&
A_\infty (k_1,k_2)=
\frac{1}{2}
\int_{-\pi/a}^{\pi/a}\frac{a^2d^2p}{(2\pi)^2}
\;\;\frac{\hat{k}^2-\hat{p}^2
-(\widehat{p+k})^2}{\hat{p}^2(\widehat{p+k})^2} \,.
}
\eeq
As usual we have introduced the lattice momenta
\[
\hat{p}_i=\frac 2a \sin \frac a2 p_i\,,
\;\;\;\;\;
\widehat{p+k}_i=\frac 2a \sin \frac a2 (p_i+k_i)\,,
\]
with
\[
p_i=\frac{2\pi n_i}{Ma}\,,
\;\;\;\;
k_i=\frac{2\pi m_i}{Ma}\,,
\;\;\;\;
n_i,m_i=0, \pm 1, \pm 2, \cdots  \pm(M/2-1), M/2
\,.
\]
for $M$ even and $|p_i|<\pi/a$ for $M\to \infty$.
Notice that for finite volume the vanishing momenta do not contribute
to the sums in \re{W2} and \re{A} and then one has an IR cutoff
\[
p_i,\,k_i\; \le \; \kir=\frac{2\pi}{Ma}
\,.
\]

\begin{figure}[ht]
\begin{center}
\mbox{\epsfig{file=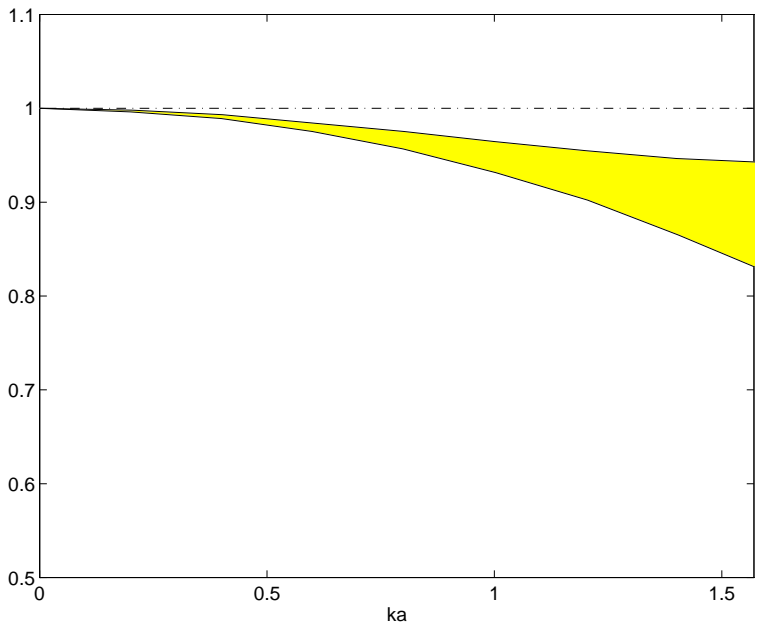,width=10.cm}}
\caption{The ratio between $A_{256}(k_1,k_2)$ 
and its leading term of Eq.~(11) plotted against $ka$; 
the spread for large $k^2$ is a lattice artifact
 representing the breaking of 
rotational invariance while the $O(k^2)$ term accounts for the curvature.
}
\label{ratio}
\end{center}
\end{figure}

We now discuss the perturbative coefficients
\beq\label{pert}
W(\al,M)= \sum_{\ell=1}\al^{\ell}\;C_\ell(M)
\,,
\;\;\;\;\;
C_\ell(M)=\frac{4\pi}{M^2}\sum_{n_1,n_2}(-4\pi A_M(k_1,k_2))^{\ell-1}
\,.
\eeq
Consider first the infinite volume case.
For small momenta $a|k_i| \ll 1$ one finds
\beq\label{soft}
A_\infty(k_1,k_2)
\simeq - \frac {1}{4\pi} \ln \left( \frac{rk^2}{\kuv^2} \right)
+ {\cal O}(a^2k^2) \,,
\;\;\;\;\;\;
r=\frac{\pi^2}{32}
\,,
\eeq
where $k^2=k_1^2+k_2^2$ (see fig.~1).

Therefore the integrand of \re{W2} at small $k_i$ is
\beq\label{gk0}
\frac{4\pi\al}{1+4\pi\al A_\infty (k_1,k_2)}
\simeq
\frac{4\pi\al}{1\,+\,b_0\,\al\,\ln(rk^2/\kuv^2)}=4\pi\al(r k^2)
\,,
\eeq
with $b_0=1$ in this case (see \re{scale}).
Therefore the integrand is given by the typical form \re{usual}
except that the running coupling is taken at the scale $rk^2$ smaller
than $k^2$.
The presence of the Landau pole is directly responsible for the
factorial behaviour at large order
\beq\label{cas}
C_\ell(\infty) \simeq b_0^{\ell-1} (\ell-1)!\, r^{-1},\quad
\ell \gg 1 \,.
\eeq
Non-leading corrections in $1/\ell$ can be estimated from \re{gk0}
by taking into account that in the soft region the integrand
$A_\infty(k_1k_2)$ approaches the running coupling at a scale $r k^2$
rather than $k^2$.

We consider now the exact evaluation of the perturbative
coefficients $C_\ell(M)$ given by \re{pert} both for $M$ finite and
infinite.

\begin{figure}[ht]
\begin{center}
\mbox{\epsfig{file=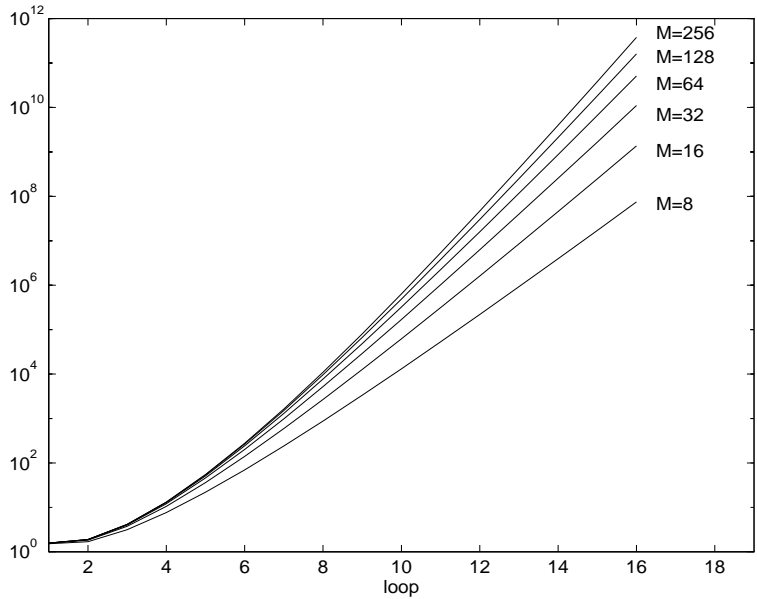,width=10.cm,height=8.cm}}
\caption{ The expansion coefficients $C_\ell$ for various 
lattice sizes $M$.
}
\label{fig:coef}
\end{center}
\end{figure}

In fig.~2 
 we report the exact coefficients $C_\ell(M)$ as function of
$\ell$ for various $M$ 
up to $M=256$; for relatively large loops ($\ell\, \gtap\, 4$) and lattice size
($M\, \gtap\, 16$) the exact coefficients $C_\ell(M)$
are fitted quite well by the
expressions
\beq\label{fit}
\bar c_\ell(M,r)=
b_0^{\ell-1} \; \int_{\kir^2}^{\kuv^2}\frac{dk^2}{k^2}\;
\left(\frac{k}{\kuv}\right)^\d\;
\ln\left(\frac{\kuv^2}{rk^2}\right)^{\ell-1}
\,,
\eeq
with $\d=2$ and $r$ given by the soft limit \re{soft} of the integrand
$r=\pi^2/32$ (see fig.~3).

\begin{figure}[t]
\begin{center}
\mbox{\epsfig{file=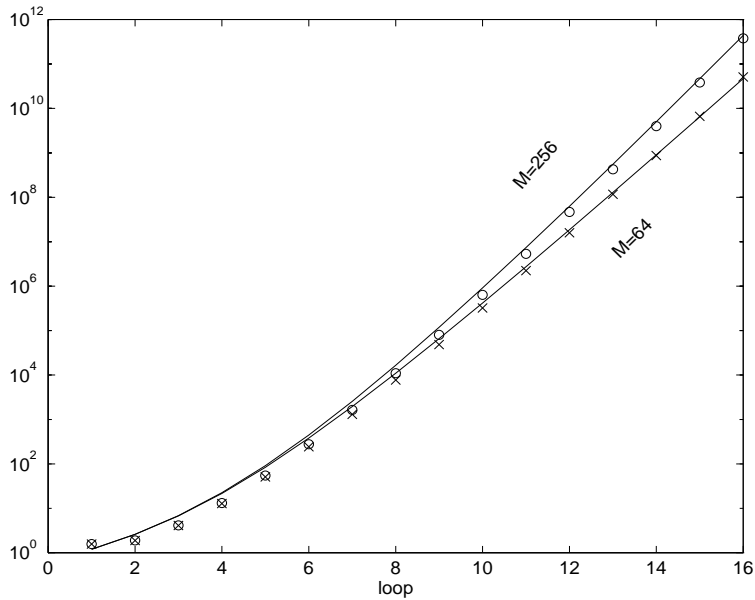,width=10.cm,height=8.cm}}
\caption{ The exact coefficients $C_\ell$  and
its approximate value given by Eq.~\rm{(14)} for $M=64$ $(+)$ and
$M=256$ $(\rm {o})$.
}
\label{fig:fit}
\end{center}
\end{figure}

At finite $M$ the factorial behaviour of the coefficients is tamed by
the IR cutoff $\kir$. To see this we plot in fig.~4 the ratio
of the exact coefficients $R_\ell(M)=C_\ell(M)/C_\ell(\infty)$ as a
function of $M$ for various $\ell$.
As expected (see the discussion based on steepest descent
estimation) this ratio vanishes by increasing $M$ at fixed $\ell$ or
by increasing $\ell$ at fixed $M$.

\begin{figure}[ht]
\begin{center}
\mbox{\epsfig{file=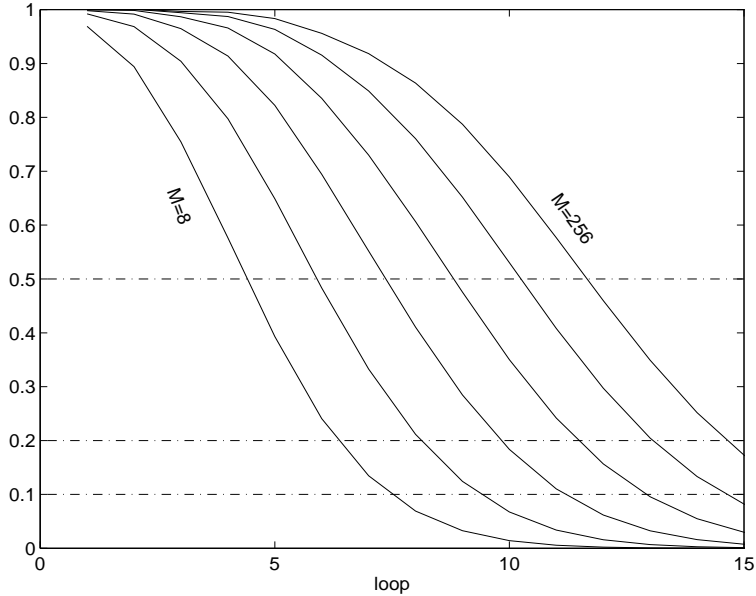,width=10.cm,height=8.cm }}
\caption{The ratios $C_\ell(M)/C_\ell(\infty)$.
}
\label{fig:ratio}
\end{center}
\end{figure}

We come now to analyze the question whether a reliable estimate of
the IR renormalon behaviour can be obtained from finite lattice
calculations. In particular we analyze the limitation \re{ellsd}
suggested by the steepest descent estimation.

\begin{figure}[t]
\begin{center}
\mbox{\epsfig{file=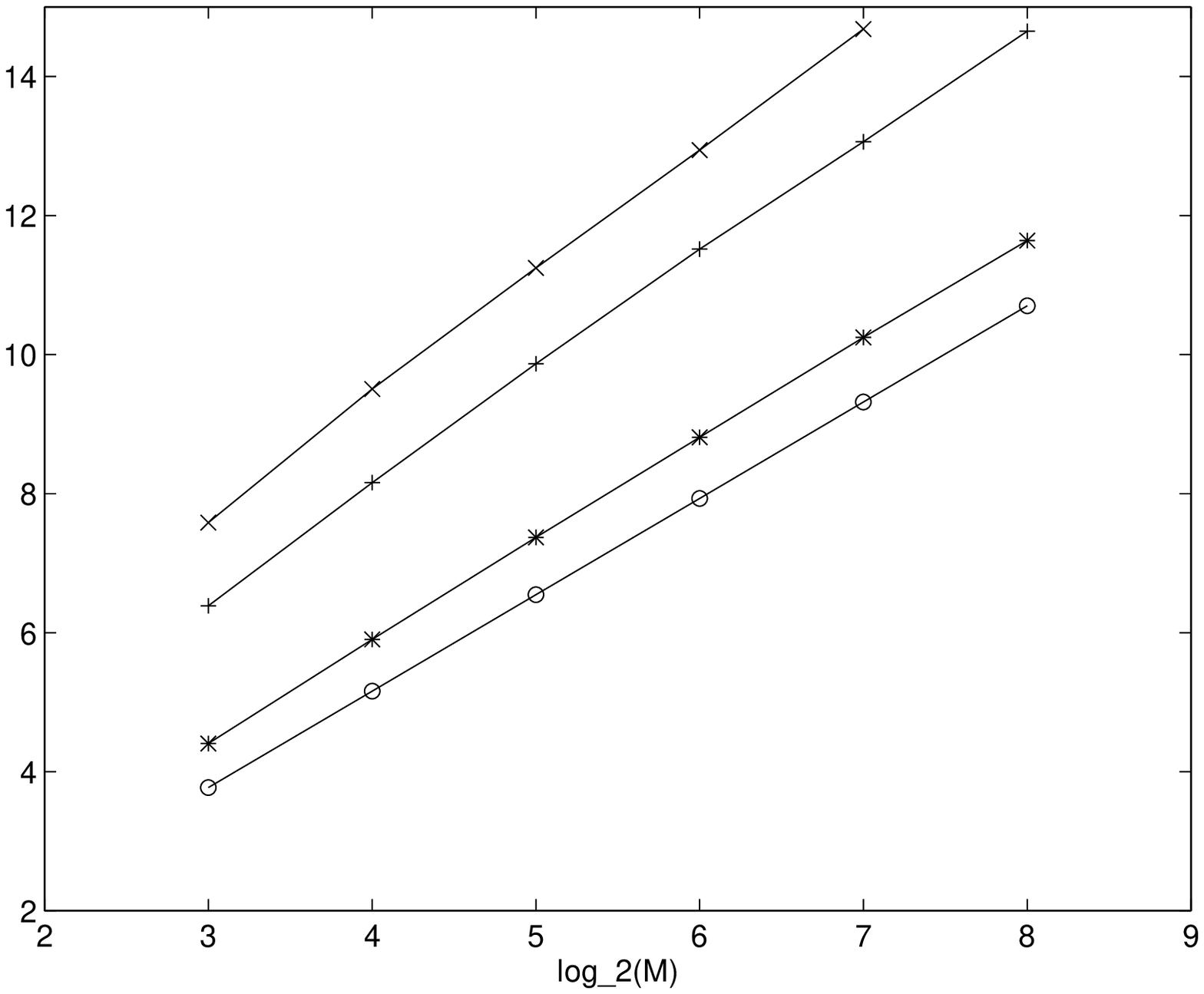,width=10.cm,height=8.cm}}
\caption{The loop order attainable with a prescribed fraction
$\sigma = .1\, (\rm{x}),\, .2 \,(+),\, .5\, (*)$ compared with the 
estimate $\ell_{sd}$ {\rm o}.
}
\label{fig:steep}
\end{center}
\end{figure}

In fig.~5 we plot $\ell_{{\rm sd}}$ as a function of $M$.
At any given $M$ we report also the maximum value of $\ell$ such
that the exact coefficient $C_\ell(M)$ is larger than a given fraction
of the infinite volume exact coefficient $C_\ell(\infty)$.
We plot the maximum $\ell$ at which $R_\ell(M) >\s$ with
$\s=0.1,\;0.2,$ and $0.5$.

The finite volume clearly tames the growth of the exact
coefficients at large $\ell$. However the factorial growth
does not stop abruptly and is still present for $\ell$
larger than the limiting value $\lsd$ given in \re{ellsd}.
For instance, for the values of $M$ considered in fig.~5,
one finds that the exact finite volume coefficients $C_\ell(M)$
are larger than $20\%$ of the infinite volume coefficient
$C_\ell(\infty)$ for all values of $\ell$ with $\ell\; \ltap \; \lsd+2$.

\section{Extension to the Yang-Mills case and conclusions}

We consider a naive extension of the previous analysis to
the gluon condensate in the Yang-Mills theory on a four dimensional
lattice.
To this end, taking into account that the expression \re{fit}
fits very well the exact coefficients in the non-linear sigma model,
we assume that also the gluon condensate coefficients
are approximated by the expression \re{fit} with $\d=4$.

\begin{figure}[ht]
\begin{center}
\mbox{\epsfig{file=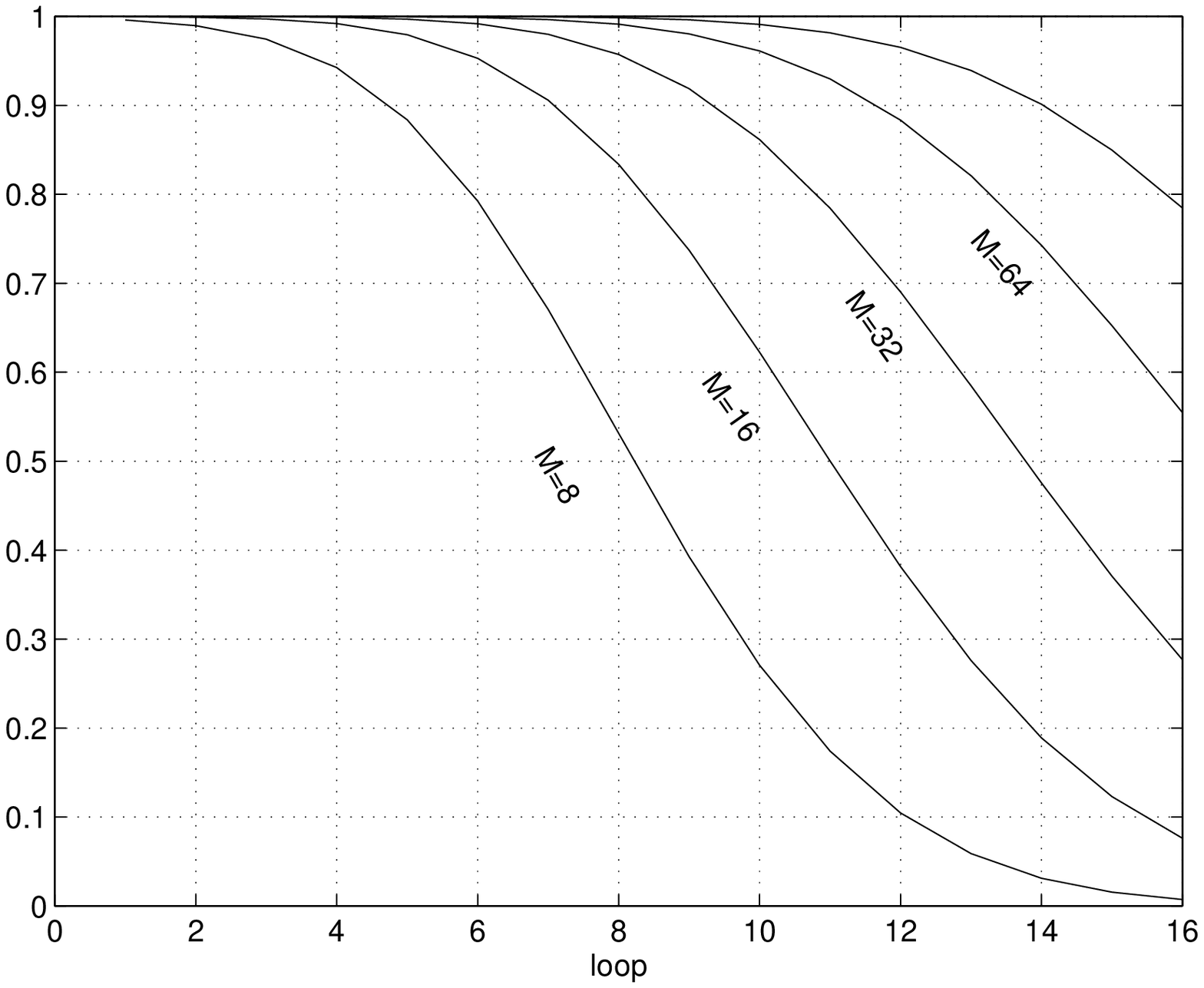,width=10.cm}}
\caption{A possible scenario for $\delta=4$ 
of the ratios $\bar c_\ell(M)/\bar c_\ell(\infty)$ as obtained
from Eq.~(14) with $\delta = 4$.
}
\label{fig:ratiofourd}
\end{center}
\end{figure}

For $r$ we assume the same value as in the non-linear sigma model.
In fig.~6 we plot  the case of the gluon condensate.
In the four dimensional case, as expected, the effect of the infrared
cutoff is less severe
and the factorial growth is still present for much larger
values of $\ell$.
For instance, in the four dimensional case the finite volume
coefficients are larger than $20\%$ of the infinite volume coefficients
for all values of $\ell$ with $\ell\; \ltap\; \lsd+4$.

We conclude that, if our fitted expression \re{fit} for the gluon
condensate perturbative coefficients is correct, the eight
loop coefficient numerically computed in \cite{DMO} is of the order of
$50\%$ of the infinite volume limit.

\vskip 1cm
\noindent
Acknowledgments
\par\noindent
We are most grateful for valuable discussions with
M.\ Beneke and G.\ Burgio.


\end{document}